 \documentclass[final,3p,times]{elsarticle}

\usepackage{amssymb}
\newcommand{\arcsinh}{\mbox{arcsinh}}

\journal{Physics Letters B}

\begin{document}

\begin{frontmatter}



\title{Polymer quantum effects on compact stars models}


\author{Guillermo Chac\'{o}n-Acosta}
\address{Departamento de
Matem\'aticas Aplicadas y Sistemas, Universidad Aut\'onoma
Metropolitana-Cuajimalpa, Vasco de Quiroga 4871, M\'exico D. F., 05348, MEXICO}
\ead{gchacon@correo.cua.uam.mx}

\author{H\'ector H. Hern\'andez}
\address{Universidad Aut\'onoma de Chihuahua, Facultad de Ingenier\'\i a, Nuevo Campus Universitario, Chihuahua 31125, MEXICO}
\address{Institute for Gravitation and the Cosmos, The Pennsylvania State University, 104 Davey Lab, University Park, Pennsylvania 16802, USA}
\ead{hhernandez@uach.mx, hhh12@psu.edu}

\begin{abstract}
In this work we study a completely degenerated fermion gas at zero temperature within a semiclassical approximation for the Hamiltonian arising in  polymer quantum mechanics. Polymer quantum systems are quantum mechanical models quantized in a similar way as in loop quantum gravity that allow the study of the discreteness of space and other features of the loop quantization in a simplified way. We obtain the polymer modified thermodynamical properties noticing that the corresponding Fermi energy is exactly the same as if one directly polymerizes  the momentum $p_F$. We also obtain the corresponding expansion of thermodynamical variables for small values of the polymer length scale $\lambda$. 
With this results we study a simple model of a compact object where the  gravitational collapse is supported by electron degeneracy pressure. We find polymer corrections to the mass of the star. When compared with typical measurements of the mass of white dwarfs we obtain a bound on the polymer length of $\lambda^2\lesssim 10^{-26}m^2$.
\end{abstract}


\end{frontmatter}



\section{Introduction}
\label{i}

Polymer quantum mechanics is the loop quantization of a quantum particle on the real line \cite{ashpqm,CVZ1}.
Various systems have been analyzed with this scheme \cite{Winkler,kun,chiou,barbero,neto,chung}, within which one can explore in an easy way certain characteristics of the theory as the discrete nature of space that leads, for instance, to modifications to the uncertainty principle \cite{ashpqm,PUR} that also have been proposed in other scenarios commonly known as GUP \cite{TGUP}, and modified dispersion relations \cite{Elias2} that may be related with the polymer mechanics \cite{maj}.

In the \textit{polymer} quantization scheme the corresponding Hilbert space $\mathcal{H}_{poly}$ is
spanned by the orthonormal base states $\{|x_j\rangle\}$, with the inner product $\langle x_i|x_j\rangle  = \delta_{i,j}$. The basic operators in this representation are the position $\hat{x}$ that acts by multiplication, and the translation operator $\hat{V}(\lambda)$ that moves the state to a distance $\lambda$, i.e. $\hat{V}(\lambda)|x_j\rangle=|x_j-\lambda\rangle$.
There is no well defined momentum operator due to the fact that translations fail to be weakly continuous in $\lambda$. Hence, a regularized  phase space function has to be constructed by the introduction of the so-called polymer length scale $\lambda$ that we interpret as a fundamental length scale \cite{ashpqm,CVZ1}.
With this regularization the polymer Hamiltonian operator can be defined as
\begin{equation}\label{ham}
  \widehat{H}_{\lambda} = \frac{\hbar^2}{2m\lambda^2} \left[ 2- \hat{V}(\lambda) -\hat{V}(-\lambda)\right]+ \hat{U}(x),
\end{equation}
where $\hat{U}(x)$ is the potential term. The action of the Hamiltonian
(\ref{ham}) decomposes the polymer Hilbert space
$\mathcal{H}_{poly}$ into a continuum of separable superselected subspaces \cite{ashpqm}.
Notice that the Hamiltonian (\ref{ham}) can be formally written as
\begin{equation}\label{ham-sin}
    \frac{\hbar^{2}}{2m\lambda^2}\widehat{\sin^{2}\Bigl(\frac{\lambda
p}{\hbar}\Bigr)} + \hat{U}(x).
\end{equation}
We can use this expression to obtain the effective Hamiltonian simply by replacing the kinetic term within the square of the sine function. This expression  contains some trace of the discreteness of the space. We will call this correspondence as \textit{polymerization}. We restrict ourselves to the one dimensional case because the effective three dimensional 
case present some complications that require numerical analysis \cite{chiapas}.

Recently the statistical mechanics of of polymer systems, was introduced in \cite{chiapas,TPQM}, where was found to have modifications to the thermodynamics induced by the polymer length scale. Since then have been studied various thermal models within this framework \cite{colima,polys,webster,polynoza}. Also a polymer Bose--Einstein condensate has been studied in which bounds for $\lambda^2$ were obtained by means of experimental data \cite{polybecs}.

In this work we explore the completely degenerated polymer Fermi gas in one dimension using a semi classical approximation of the polymer Hamiltonian. The completely degenerated Fermi gas at vanishing temperature is the starting point for the study of compact stars models where the degeneracy pressure must balance the gravitational collapse. We look at the corresponding polymer corrections on the thermodynamical variables. Particularly the pressure, which gives us a small modification in the mass of the compact object, allowing us to bound the polymer length through typical measurements of the mass of white dwarfs.

\section{One dimensional Fermi gas at zero temperature}
\label{1}

Let us begin with a high density, one dimensional gas at zero temperature, constituted by particles obeying the Fermi-Dirac statistics. We remember the calculation of the thermodynamic expressions for this system, since the one-dimensional, differs from the three-dimensional one reported in any textbook. Fermions populate the states of lowest energy and, due to the Pauli exclusion principle, each state will be occupied by only one fermion and $g_s=2s+1$ fermions for each energy level. The energy levels will be filled until the Fermi energy $E_F$ is reached. The Fermi energy is defined as the highest energy state occupied at zero temperature $E_F= \mu(T=0)$, where $\mu$ is the chemical potential.
For $T=0$ the occupancy number is very accurately described by a Heaviside function, as can be seen by performing the limit $T\rightarrow 0\,$ of the Fermi distribution
\begin{equation}\label{lim}
    \lim_{T\rightarrow 0} n_{_{FD}}(E)=\Theta(\mu-E)=\left\{
                                                   \begin{array}{ll}
                                                     1, & \hbox{$E\leq\mu$;} \\
                                                     0, & \hbox{$E>\mu$.}
                                                   \end{array}
                                                 \right.
\end{equation}
In this way one can compute the thermodynamic quantities like the particle number and the internal energy as follows, \cite{greiner}
\begin{eqnarray}
N & = & \int^{\infty}_{0}g(E)\Theta(\mu-E)dE, \label{N} \\
U & = & \int^{\infty}_{0}g(E)\Theta(\mu-E)EdE, \label{U} 
\end{eqnarray}
where $g(E)$ is the state density (or energy level density) obtained by deriving the number of configurations in phase space, or phase volume $\Sigma$, i.e. $g(E) = \frac{d\Sigma}{dE}$.
For a one dimensional fermion gas $g(E)=\frac{Lg_s}{2\pi \hbar} \sqrt{\frac{m}{2}} E^{-1/2}$, where $L$ is the length of the box and $m$ the mass of the particles. Performing the integrals of (\ref{N}) and (\ref{U}), the total number of particles and the internal energy are
\begin{eqnarray}
N & = & \frac{Lg_s}{2\pi \hbar} \sqrt{2mE_F}, \label{N0} \\
U & = & \frac{Lg_s}{12\pi \hbar m} \left(2mE_F\right)^{3/2}, \label{U0}
\end{eqnarray}
where $E_F$ is the Fermi energy that depends on the dimension and the particle density $\frac{N}{L}$, \cite{greiner}. In this case is
\begin{equation}\label{EF}
    E_F=\frac{1}{2m}\left( \frac{N}{L} \frac{2\pi \hbar}{g_s} \right)^2.
\end{equation}
Using (\ref{N0}) and  (\ref{U0}) it is possible to obtain the energy per particle in terms of $E_F$
\begin{equation}\label{UN}
   \frac{U}{N}=\frac{1}{3}E_F.
\end{equation}
The pressure can be calculated from the logarithm of the grand canonical partition function \cite{greiner}, such that in one dimension and in the zero temperature regime it turns to be
\begin{equation}\label{P}
   P= \frac{g_s}{2 \pi \hbar} \int_0^{p_F}p\,dp \frac{dE}{dp},
\end{equation}
in the non relativistic case where $p=\sqrt{2mE}$, the pressure is 
\begin{equation}\label{P0}
   P= \frac{g_s}{6 \pi \hbar m}p_F^3 = \frac{g_s}{6 \pi \hbar m}(2mE_F)^{3/2} =2\frac{U}{L},
\end{equation}
so we can write the pressure in terms of Fermi energy as
\begin{equation}\label{PEF}
   P=\frac{2}{3}\frac{N}{L}E_F.
\end{equation}

On the oner hand, we want to have the thermodynamic properties of Fermi gas in the ultra relativistic case. The dispersion relation is just $E=cp$, with $c$ the speed of light. Thus, when we integrate (\ref{U}) in the momentum space, we obtain the following
\begin{equation}
\label{Uur}
  U_{UR} = \frac{Lg_sc}{4\pi \hbar} p_F^2.
\end{equation}
And for (\ref{P}) we obtain
\begin{equation}\label{Pur}
   P_{UR}= \frac{g_s c}{4 \pi \hbar}p_F^2  = \frac{U_{UR}}{L}.
\end{equation}

Indeed, we can write the full relativistic case by considering the dispersion relation corresponding to the relativistic kinetic energy as $E = mc^2\left(\sqrt{1+\frac{p^2}{m^2c^2}}-1\right)$ in all previous expressions. In such a case we obtain the following for the energy (\ref{U}) and the pressure (\ref{P})
\begin{eqnarray}
U & = & \frac{ L g_s}{2 \pi \hbar} \int_0^{p_F} mc^2\left(\sqrt{1+\frac{p^2}{m^2c^2}}-1\right) \,dp =   \frac{L g_s m^2c^3}{4 \pi \hbar}\left[ \frac{p_F}{mc}\sqrt{1+\frac{p_F^2}{m^2c^2}}  +  \arcsinh \left( \frac{p_F}{mc} \right)  -\frac{2p_F}{mc}  \right], \label{Ur} \\
P & = & \frac{g_s c}{2 \pi \hbar} \int_0^{p_F} \frac{p/mc}{\sqrt{1+\frac{p^2}{m^2c^2}}}p\,dp = \frac{g_s m^2c^3}{4 \pi \hbar}\left[ \frac{p_F}{mc}\sqrt{1+\frac{p_F^2}{m^2c^2}} -  \arcsinh \left( \frac{p_F}{mc} \right)  \right].    \label{Pr}
\end{eqnarray}
When we perform the corresponding expansions in non-relativistic $p_F \ll mc$ and ultra-relativistic $p_F\gg mc$, cases, we recover what has already obtained before. Notice that in both relativistic and ultra relativistic cases the number of particles does not change. 

Our main interest in cold Fermi gases, is that we would like to study compact stars such as white dwarfs, as mentioned before in particular with the effective polymer model, which we introduce in the next section.

A standard white dwarf can be modeled as a gas of mass $M$, made up by $N$ degenerate electrons whose pressure balance the gravitational collapse. Also in the one-dimensional case a compact object is formed since the electron degeneracy pressure increases faster than the gravitational pressure, which leads to a stable situation. However, as the length-mass relationship is not the same as in the three-dimensional case where the radius decreases as the mass increases, one can not say that this object is precisely a white dwarf. In such a case the gravitational pressure in thermodynamic equilibrium is
\begin{equation}
\label{Pstar}
  P(L) = \frac{kM^2}{L^2},
\end{equation}
where $k$ is a constant related with the gravitational constant and with a factor of order unity that is related with the corresponding density profile of the system. The corresponding gravitational potential in the polymer case would be similar to that considered in \cite{colima} for the Coulomb potential. This expression can be evaluated for non relativistic electrons with $g_s=2$, with pressure (\ref{P0}), whose main contribution to the mass comes from its nuclei, that is $M=2Nm_n$, with $m_n$ stands for the mass of a nucleon. Then, Eq. (\ref{Pstar}) leads to
\begin{equation}
\label{1DLM}
  L = \frac{M}{\rho_*}.
\end{equation}
where $\rho_*=\frac{24 m m_n^3 k}{\pi^2 \hbar^2}$ is the critical mass density of the model.
In the polymer quantization scheme the thermodynamic quantities are modified, so let us look at how these changes influence the thermodynamics of the gas of fermions.

\section{Polymer Fermi gas at zero temperature}
\label{pfg}

In order to analyze the posible effects induced by the polymer length scale we use a semiclassical approximation for the polymer Hamiltonian as proposed in \cite{chiapas, polybecs,vallarta}
\begin{equation}\label{Ham}
E=\frac{\hbar^{2}}{2m\lambda^2}\sin^{2}\Bigl(\frac{\lambda
p}{\hbar}\Bigr).
\end{equation}
We work with the standard Fermi-Dirac distribution, however, as obtained in \cite{TPQM} for the Planck distribution,
 there could be modifications to the former due to the polymer quantization scheme. By considering valid approximation (\ref{lim}) for the distribution, we now compute the level density obtained from the
 number of configurations $\Sigma_{_{poly}}$ that has polymer corrections due to (\ref{Ham})
\begin{equation}\label{config-poly}
    \Sigma_{_{poly}} =  \frac{Lg_s}{2\pi \hbar} \int dp = \frac{Lg_s}{4\pi \lambda} \int \frac{dy}{\sqrt{y}\sqrt{1-y}},
\end{equation}
where we introduce a dimensionless variable $y$ and we call $E_{_{poly}}$ the \textit{polymer energy}
\begin{equation}\label{E-poly}
    y \equiv \frac{E}{E_{_{poly}}},\;\;E_{_{poly}} \equiv \frac{\hbar^2}{2m\lambda^2}.
\end{equation}
It follows that the polymer state density is
\begin{equation}\label{g-poly}
    g_{_{poly}}(E) = \frac{Lg_s}{4\pi \lambda E_{_{poly}}} \left[ \frac{E}{E_{_{poly}}} \left( 1- \frac{E}{E_{_{poly}}} \right)\right]^{-1/2}.
\end{equation}

Using the same statistical definitions for the thermodynamic quantities we can integrate (\ref{N}) using the modified state density $g_{_{poly}}$ of (\ref{g-poly}). Therefore, the particle number is
\begin{equation}\label{N-poly}
    N = \frac{Lg_s}{4\pi \lambda } \int_0^{E_F/E_{_{poly}}}\,\frac{dy}{\sqrt{y}\sqrt{1-y}} = \frac{Lg_s}{2\pi \lambda}\arcsin\left(\sqrt{\frac{E_F}{E_{_{poly}}}}\right),
\end{equation}
We should mention that the integral in the previous expression is valid only when $0<E_F<E_{_{poly}}$. The polymer Fermi energy is bounded from above, polymerization induces an energy cutoff, \cite{CVZ1}. This can be seen more clearly if we write an expression for the polymer Fermi energy in terms of particle density from (\ref{N-poly})
\begin{equation}\label{EF-poly}
    E_F = \frac{\hbar^2}{2m\lambda^2}\sin^2\left( \frac{N}{L} \frac{2\pi\lambda}{g_s} \right),
\end{equation}
which is equivalent to replace $p_F^2\rightarrow \sin^2(\lambda p_F/\hbar)\hbar^2\lambda^{-2}$ in (\ref{EF}), with $p_F=\frac{N}{L}\frac{2\pi \hbar}{g_s}$. Also notice that in the limit when $\lambda \rightarrow 0, \; \Rightarrow \; E_{_{poly}} \rightarrow \infty$ and $E_F$ has no longer an upper bound,
\begin{equation}\label{EF-poly-corr}
    E_F \approx \frac{2\pi^2 \hbar^2}{g_s^2 m} \left( \frac{N}{L} \right)^2 -\lambda^2  \frac{8\pi^4 \hbar^2}{3g_s^4 m} \left( \frac{N}{L} \right)^4 + \mathcal{O}(\lambda^4).
\end{equation}

We can also expand (\ref{N-poly}) for small values of polymer length scale, then to first order we recover (\ref{N0}) for $N$, but we also have corrections up to second order in $\lambda^2$
\begin{equation}\label{N-poly-corr}
    N \approx \frac{Lg_s}{2\pi \hbar} \sqrt{2mE_F}  + \lambda^2 \frac{Lg_s}{12\pi m \hbar^3}(2mE_F)^{3/2} + \mathcal{O}(\lambda^4).
\end{equation}

The corresponding internal energy can be calculated from (\ref{U}) and (\ref{g-poly}) and it results
\begin{eqnarray}
  U &=& \frac{Lg_s}{4\pi \lambda} E_{_{poly}} \int_0^{E_F/E_{_{poly}}}\,\frac{\sqrt{y}}{\sqrt{1-y}}\,dy \nonumber \\
   &=& \frac{Lg_s}{4\pi \lambda} \frac{\hbar^2}{2m\lambda^2} \left[\arcsin\left(\sqrt{\frac{E_F}{E_{_{poly}}}}\right) - \sqrt{\frac{E_F}{E_{_{poly}}}\left(1-\frac{E_F}{E_{_{poly}}} \right) } \right], \label{U-poly}
\end{eqnarray}
which, in the limit of small polymer length, one recovers (\ref{U0}) at first order for $U$
\begin{equation}\label{U-poly-corr}
    U \approx \frac{Lg_s}{12\pi \hbar m} \left(2mE_F\right)^{3/2}  + \lambda^2 \frac{Lg_s}{40\pi m \hbar^3}(2mE_F)^{5/2} + \mathcal{O}(\lambda^4).
\end{equation}
It is also possible to obtain expressions for the energy per particle and its corresponding expansion in $\lambda$
\begin{eqnarray}
  \frac{U}{N} &=& \frac{\hbar^2}{4m\lambda^2}\left[ 1- \frac{\sqrt{\frac{E_F}{E_{_{poly}}}\left(1-\frac{E_F}{E_{_{poly}}}  \right) }}{\arcsin\left(\sqrt{\frac{E_F}{E_{_{poly}}}}\right)} \right], \\
   \frac{U}{N} &\approx& \frac{E_F}{3} + \lambda^2  \frac{4m}{45\hbar^2}E_F^2 + \mathcal{O}(\lambda^4).
\end{eqnarray}

The pressure can be calculated from (\ref{P}) and (\ref{Ham}), giving
\begin{equation}\label{P-poly}
    P = \frac{g_s \hbar^2}{8m \pi \lambda^3}  \Biggl[\left(2\frac{E_F}{E_{_{poly}}}-1\right)\arcsin\left(\sqrt{\frac{E_F}{E_{_{poly}}}}\right) + \sqrt{\frac{E_F}{E_{_{poly}}}\left(1-\frac{E_F}{E_{_{poly}}} \right) } \Biggr],
\end{equation}
therefore, the relation (\ref{P0}) is just satisfied for the leading order of the expansion for small values of the polymer length 
\begin{equation}\label{P-poly-corr}
    P \approx \frac{g_s}{6\pi \hbar m} \left(2mE_F\right)^{3/2}  + \lambda^2 \frac{g_s}{60\pi m \hbar^3}(2mE_F)^{5/2} + \mathcal{O}(\lambda^4),
\end{equation}
notice that the second order of (\ref{P-poly-corr}) differs form (\ref{U-poly-corr}) for a factor of $2/3$.

Interestingly corrections in all the above expressions are always of second order in the polymer length, as already noted in \cite{chiapas,TPQM,colima, polybecs}.

For the ultra relativistic case we can consider as a simple  dispersion relation for polymer relativistic particles, that 
\begin{equation}
\label{EpolyUR}
  E = \frac{c \hbar}{\lambda}\sin\left(\frac{\lambda p}{\hbar}\right).
\end{equation}
Thus, we can integrate (\ref{U}) in the momentum space using (\ref{EpolyUR}) obtaining the following
\begin{equation}
\label{poly-Uur}
  U_{UR} = \frac{Lg_sc \hbar}{2\pi \lambda^2} \left[ 1 - \cos\left(\frac{\lambda p_F}{\hbar} \right)  \right],
\end{equation}
we can expand this expression for small $\lambda$
\begin{equation}
\label{poly-Uur-corr}
  U_{UR} \approx \frac{Lg_sc}{4\pi \hbar} p_F^2 - \lambda^2 \frac{Lg_sc}{48\pi \hbar^3}p_F^4 + \mathcal{O}(\lambda^4).
\end{equation}
We can also calculate the pressure 
\begin{eqnarray}
  P_{UR} &=& \frac{g_s c \hbar}{4 \pi \lambda^2}\sin\left(\frac{\lambda p_F}{\hbar}\right)^2,  \label{poly-Pur}\\
   P_{UR} &\approx& \frac{g_s c}{4 \pi \hbar}p_F^2 - \lambda^2  \frac{g_s c}{12 \pi \hbar^3}p_F^4 + \mathcal{O}(\lambda^4). \label{poly-Pur-corr}
\end{eqnarray}
again, the second order in $\lambda$ of the expansion of $P_{UR}$ differs from second order of (\ref{poly-Uur-corr}) by a factor of $1/4$, with respect to the standard relation (\ref{Pur}).

The polymer fully relativistic case is quite complicated when performing the integrals. Since we are interested in are the limiting cases, we leave that case to be reported elsewhere. Now, we investigate how the polymer corrections influence in models of compact stars .

\section{Polymer compact star model}%
\label{stars}

Let us consider the compact object model introduced at the end of section \ref{1}, but with the polymer modifications to fermion's thermodynamics. We notice that the degeneracy pressure of electrons is changed by polymer effects (\ref{P-poly}) or (\ref{P-poly-corr}), so this would modify the mass of the compact object. In order to maintain the electron gas at a given density, the electron degeneracy pressure must resist the gravitational collapse. Therefore, it is necessary to balance (\ref{Pstar}) with (\ref{P-poly}) or (\ref{P-poly-corr}). If we use (\ref{P-poly}), we notice that as $E_F$ also has a polymer expression, we have a transcendental equation for the mass. However, using (\ref{P-poly-corr}) and (\ref{EF-poly-corr}), we can find the second order polymer corrections to the mass, so
\begin{equation}
\label{poly-star-mass}
  M = \rho_* L\left( 1+  \lambda^2 \frac{2 \pi^2}{5}\frac{N^2}{L^2} + \ldots \right),
\end{equation}
where clearly $\rho_* L$ is the mass of the object without polymer corrections according to (\ref{1DLM}). Typically, for a white dwarf star the inverse of the average distance is of the order of $10^{12}m^{-1}$, so take this value to estimate the change in mass. For polymer length there are an estimate obtained from Bose-Einstein condensates experiments \cite{polybecs}, which is $\lambda^2\lesssim10^{-16}m^2$. Nevertheless, when we use that value on (\ref{poly-star-mass}) we obtain that the polymer modification is of order $10^8$, which is not possible. However, we can improve the bound for $\lambda^2$, using typical measurements of the mass for a white dwarf. For example, in \cite{WD} a mass of $M/M_{\odot} = 0.529 \pm 0.01$ was obtained,  bringing a value of $\lambda^2\lesssim10^{-26}m^2$, which is lower than that obtained previously.

There also exist another configuration where the object is composed mainly by ultra relativistic electrons, so its mass is well approximated by its internal energy $M=U_{UR}/c^2$. The electronic degeneracy pressure again resist the gravitational collapse so Eq. (\ref{Pstar}) is also fulfilled. In equilibrium is possible to calculate the size of the compact object with polymer corrections, so using (\ref{poly-Uur-corr}) and (\ref{poly-Pur-corr}) 
we find
\begin{equation}
\label{poly-star-mass}
  L \approx \frac{R_S}{2}\left( 1+  \lambda^2 \frac{p_F^2}{4\hbar^3} + \ldots \right),
\end{equation}
where $R_S=2kM/c^2=2kU_{UR}/c^4$ is the corresponding Schwarzschild radius, so we keep the symbol. Although the changes are of the same order in $\lambda^2$, we must not forget that there are also polymer corrections to $p_F$ and to $R_S$ through $U_{UR}$.

\section{Conclusions}%
\label{c}

Polymer quantum mechanics is a quantum mechanical model in which certain aspects of loop quantization, such as the minimum length scale of discrete space, can be studied \cite{ashpqm,CVZ1}. In particular a thermostatistical description of this type of quantization has been studied for solids, ideal gases \cite{TPQM}, hydrogen gases \cite{colima}, and some other thermal systems\cite{colima,polys,webster,polynoza}. In most cases corrections  induced by $\lambda$ to the thermodynamic properties such as pressure, entropy, energy and heat capacity are found. For Bose-Einstein gases, were found bounds for  $\lambda$ coming for typical experimental conditions \cite{polybecs}. This type of behavior has also been found in different approaches such as generalized uncertainty principle (GUP) \cite{Husain-Mann, newgup}.

In this paper we calculate the thermodynamical properties for a one dimensional polymer Fermi gas within a semi classical approximation in the polymer Hamiltonian. We considered the Fermi--Dirac distribution for the $t=0$ case, with no polymer modifications.

The corresponding total number of particles, internal energy and pressure, and the Fermi energy have modifications related with the polymer length scale. We found that in an expansion for small Ápolymer length, the first corrections are of order $\lambda^2$. The polymer Fermi energy is the same as if we polimerized directly the Fermi momentum $p_F$. It is also worth mentioning that this polymer thermal quantities are defined just for a Fermi energy grater than zero and less than $E_{_{poly}}$, that is inversely proportional to $\lambda$, it induces a cut off in the energy.

We also investigate polymer effects on the thermal properties of compact objects models. We found that polymer corrections modify the effective mass and size of the star. Indeed we found that for keeping these effects small enough, the value of the polymer length should remain $\lambda^2\lesssim 10^{-26}m^2$, when compared with white dwarfs typical measurements. This bound improves the bound that has been obtained with other experiments. 
Unlike the work in GUP, where theoretical corrections to the properties of white dwarfs are obtained \cite{JHEP, IJMPD}, in the polymer case it was possible to set a bound comparing with typical observations. In fact if we consider the corrections found in those works for bounding the polymer scale, it follows that $\lambda$ would be even smaller. 

It is expected to have similar corrections in the 3D case. It has been suggested that under strong enough magnetic fields, the quantization becomes anomalous, which induces a magnetic transverse collapse of the gas, and then it is possible that the Fermi-Dirac gas reduces to a one-dimensional system\cite{lofer}. So it is possible that, by introducing such effects, the one-dimensional system become of interest and may induce further polymer effects.

\section*{Acknowledgments}%
\label{ak}

Partial support from the grant CONACYT CB-2008-01/101774 is
acknowledged. H.H. was supported by CONACYT sabbatical grant  C000/826/13.





\end{document}